\documentclass[a4paper,11pt]{article}
\pdfoutput=1

\usepackage{amssymb,amsmath,bm}
\usepackage{a4wide}
\usepackage{color}
\usepackage{slashed}
\usepackage{graphicx}
\usepackage[latin1]{inputenc}
\usepackage{amsfonts}
\usepackage{lscape}
\usepackage{amsthm}
\usepackage{float}
\usepackage{booktabs}
\usepackage{array}
\usepackage{rotating}
\usepackage[numbers, sort]{natbib}
\usepackage{multirow}

\usepackage[small,bf]{caption}
\newcommand{\eps}{\epsilon}
\newcommand{\ord}[1]{\mathcal{O}\left( #1 \right)}

\newcommand{\Qb}{\overline{Q}}
\newcommand{\Ub}{\overline{U}}
\newcommand{\Db}{\overline{D}}

\newcommand{\Hc}{\mathcal{H}}
\newcommand{\Hcb}{\overline{H}}
\newcommand{\Sb}{\overline{S}}

\newcommand{\five}{\mathbf{5}}

\begin{document}

\vspace{1cm}
\begin{titlepage}
\vspace*{-1.0truecm}
\begin{flushright}
MPP-2012-60 \\
TUM-HEP-829/12
\end{flushright}

\vspace{0.4truecm}

\begin{center}
\boldmath

{\Large\textbf{The Messenger Sector of SUSY Flavour Models \\ and Radiative Breaking of Flavour Universality}}			

\unboldmath
\end{center}

\vspace{0.4truecm}

\begin{center}
{\bf Lorenzo Calibbi$^a$, Zygmunt Lalak$^{b,c}$,\\
Stefan Pokorski$^{b,d}$, Robert Ziegler$^{d,e}$
}
\vspace{0.4truecm}

{\footnotesize

$^a${\sl Max-Planck-Institut f\"ur Physik (Werner-Heisenberg-Institut),
 F\"ohringer Ring 6, \\ D-80805 M\"unchen, Germany}\vspace{0.2truecm}

$^b${\sl Institute of Theoretical Physics, Faculty of Physics, University of Warsaw, Ho\.za 69,\\ 00-681, Warsaw, Poland}\vspace{0.2truecm}

$^c${\sl CERN Physics Department, Theory Division,\\ CH-1211 Geneva 23, Switzerland \vspace{0.2truecm}}

$^d${\sl TUM-IAS, Technische Universit\"at M\"unchen,  Lichtenbergstr.~2A,\\ D-85748 Garching, Germany \vspace{0.2truecm}}

$^e${\sl Physik Department, Technische Universit\"at M\"unchen,
James-Franck-Stra{\ss}e, \\D-85748 Garching, Germany}

}
\end{center}

\vspace{0.4cm}
\begin{abstract}
\noindent The flavour messenger sectors and their impact on the soft SUSY breaking terms are investigated in SUSY flavour models. In the case when the flavour scale $M$ is below the SUSY breaking mediation scale $M_S$, the universality of soft terms, even if assumed at $M_S$, is radiatively broken. We estimate this effect in a broad class of models. In the CKM basis that effect gives flavour off-diagonal soft masses comparable to the tree-level estimate based on the flavour symmetry.   
\end{abstract}
\end{titlepage}

\section{Introduction}

A common approach to explain the observed hierarchies in fermion masses and mixings is in terms of flavour symmetries. 
The matter fields transform under the flavour symmetry, which is 
spontaneously broken by the vacuum expectation values (vevs) of scalar fields that we will call flavons in the following.
Most Yukawa couplings are forbidden at the renormalisable level and only arise from higher-dimensional operators involving suitable powers of the flavons as determined by the flavour symmetry \cite{FN}. The flavour hierarchies are then explained by small order parameters given by the ratio of flavon vev and the UV cutoff scale. This scale itself remains undetermined and can in principle be as large as the Planck scale. In case it is smaller, one can interpret this cutoff scale as the typical mass scale of new degrees of freedom, the so-called ``flavour messengers''. The dynamics of this sector may have important impact on low-energy physics even if its characteristic scale is very high. In this work we want to systematically analyse the structure of the messenger sector and show that it can indeed have important consequences both for Yukawa couplings and soft masses in the MSSM.   
 
In the UV completion that contains the messenger sector, small fermion masses can be thought to arise from small mixing of light fields either with heavy fermions or with heavy scalars. In the first case, which has been studied extensively in Refs. \cite{MMM1,MMM2}, fermion masses correspond to three light eigenvalues of a large mass matrix. In the second case, light masses arise from small vevs of the heavy scalars. This latter case has received less attention despite its structure is much simpler. Such scalar messengers are very suitable to generate texture zeros in the Yukawa matrix \cite{quilt}, that is, the vanishing of certain entries although allowed by the flavour symmetry.\footnote{More recently, some phenomenological consequences of the messenger sector have been discussed in \cite{Kadota:2010cz,Luca} and in \cite{Antusch:2010es} where also heavy
scalars have been employed.} 

Independently of the kind of messengers, in general one needs many of them. Since these fields carry SM quantum numbers, they contribute to the RG evolution of SM gauge couplings. The requirement of maintaining perturbativity up to certain scale puts a lower bound on the messenger masses as a function of their number. In this work we require perturbative physics up to a very high scale of supersymmetry breaking mediation, the Planck scale for Gravity Mediation or the Gauge Mediation scale. In consequence the flavour messengers must be very heavy. Direct effects of their exchange are then strongly suppressed\footnote{In models with a low fundamental cutoff one can take these fields down to low scale where they can give rise to large flavour-changing effects which make such scenarios testable at experiments \cite{CLPZlow}.}, but, when the flavour messenger scale is lower than the SUSY breaking mediation scale, they are relevant for the flavour dependence of the soft SUSY breaking terms. This is because in the supersymmetric theories we are considering the soft SUSY breaking terms are sensitive to any dynamics that couple to light fields. 

One can distinguish tree-level effects on the soft mass terms, generated by integrating out the flavour messengers and determined by the flavour symmetry alone~\cite{LPR}, and radiative effects summarised by the RG evolution between the SUSY breaking mediation scale $M_S$ and flavour messenger scale $M$, with the flavour messengers included. The latter are particularly interesting as they necessarily break flavour universality of soft terms, even if it is assumed (or naturally present) as the initial condition at $M_S$. We quantitatively estimate these effects and show that they always give in the super-CKM basis flavour off-diagonal terms in soft masses comparable (in their order of magnitude) to those one would obtain at tree-level without assuming universality. Thus the evolution of the soft terms according to the MFV hypothesis~\cite{MFV} from their initial universal values is possible only if $M$ is larger than $M_S$. For Gravity Mediation, the flavour physics is then pushed to the Planck scale. 

The structure of our paper is as follows: In Section 2 we first discuss the general structure of the messenger sector, in particular the two possibilities of UV completions with heavy fermions and heavy scalars, respectively. This allows us to estimate the typical number of messengers that we use in Section 3 to calculate the bounds on the messenger scale from perturbativity. In Section 4 we discuss the impact of the messenger sector on the flavour structure of soft terms. In particular we present constraints on light rotation angles valid in large classes of flavour models. We finally conclude in Section 5. In the Appendix we provide explicit examples of UV completions of three popular flavour models in the literature in order to illustrate our general discussion.

\section{The Structure of the Messenger Sector}

\label{FN-UVC}

We want to consider models with a general flavour symmetry group $G_F$ spontaneously broken by the vevs of 
the flavon superfields $\phi_I$. The MSSM Yukawa couplings arise from higher-dimensional $G_F$ invariant operators involving the flavons 
\begin{align}
W_{yuk} & = y_{ij}^U q_i u^c_j h_u + y_{ij}^D q_i d^c_j  h_d &  y_{ij}^{U,D} & \sim \prod_{I} \left( \frac{\langle\phi_I\rangle}{M} \right)^{n^{U,D}_{I,ij}}, 
\label{yuk}
\end{align}
where the suppression scale $M \gtrsim \langle\phi_I\rangle $ is the typical scale of the flavour sector dynamics. The coefficients of the effective operators are assumed to be $\ord{1}$, so that Yukawa hierarchies arise exclusively from the small order parameters $\epsilon_I \equiv {\langle\phi_I\rangle}/{M}$. The transformation properties of the MSSM fields and the flavons under $G_F$ are properly chosen, so that $\epsilon_I$ together with their exponents $n^{U,D}_{I,ij}$ reproduce the observed hierarchies of fermion masses and mixings.

\begin{figure}[h]
\centering
\includegraphics[width=0.9\textwidth]{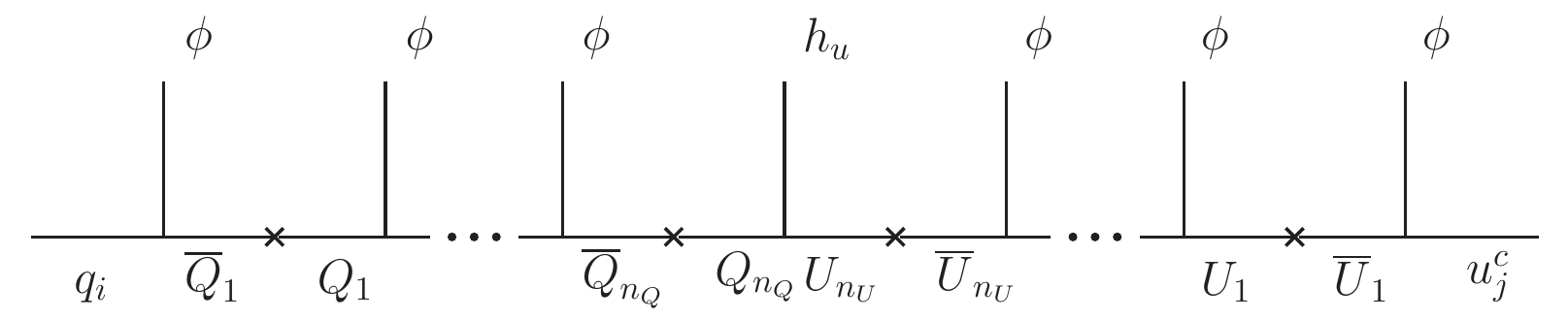}
\caption{\label{Fchain} Schematic supergraph for Fermion UVC.}
\end{figure}

In order to UV-complete these models one has to "integrate in" heavy fields at the scale $M$. These messenger fields are vectorlike and charged under $G_F$. They couple to the flavons and the light matter. In the fundamental theory they mix either with MSSM matter or Higgs fields (after flavour symmetry breaking). In the first case one has to introduce chiral superfields $(Q + \overline{Q},~U + \overline{U},~D + \overline{D})$ with the quantum numbers of the MSSM matter fields (see Fig.~\ref{Fchain}). 

\begin{figure}[h]
\centering
\includegraphics[width=0.92\textwidth]{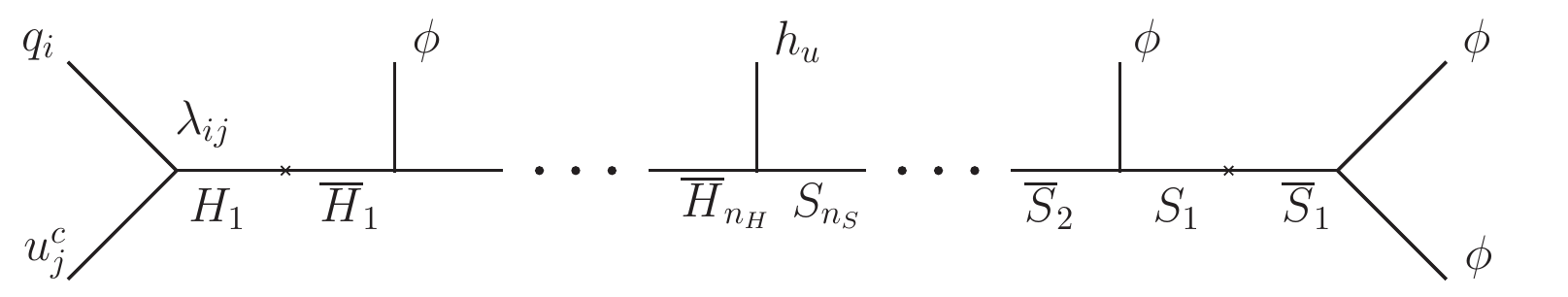}
\caption{\label{Hchain}Schematic supergraph for Higgs UVC.}
\end{figure}

\noindent In the second case one introduces chiral fields $(H + \overline{H},~S + \overline{S})$ with the quantum numbers of the MSSM Higgs fields and $R_P$-even gauge singlets, respectively (see Fig.~\ref{Hchain}). 

\noindent We are now going to discuss in more detail these two possibilities, to which we refer as ``Fermion UV completion" (FUVC) and ``Higgs UV completion" (HUVC). Although we restrict to the pure cases, a UV completion involving both kind of fields is also viable. 
Such a case is a straightforward generalisation of the following discussion. 

\subsection{UV completion with heavy fermions}
In order to UV-complete the effective theory above with fermionic messengers, one has to introduce certain numbers of vector-like superfields with light fermion quantum numbers
\begin{align*}
 Q_\alpha + \Qb_\alpha \qquad \qquad  U_\alpha + \Ub_\alpha \qquad \qquad  D_\alpha   +\Db_\alpha ,
\end{align*}
with superpotential interactions of the schematic form 
\begin{align}
\label{flavonint}
W & \supset M_Q \Qb_\alpha Q_\alpha + M_U \Ub_\alpha U_\alpha + \phi_I \left( \Qb_\alpha Q_\beta + \Ub_\alpha U_\beta \right) \nonumber \\ 
& + \phi_I \left(\Qb_\alpha q_i + \Ub_\alpha u^c_i \right)+ h_u \left( Q_\alpha U_\beta + Q_\alpha u^c_i + q_i U_\alpha \right),       
\end{align}
where we restricted to the up-sector for simplicity. Moreover we have dropped all couplings that are assumed to be $\ord{1}$, since all hierarchies should arise from the symmetry breaking alone (which also implies $M_Q \sim M_U \sim M$). 

The allowed couplings follow directly from the transformation properties of the messengers under $G_F$. They have to be chosen appropriately, such that the effective Yukawa couplings in Eq.~(\ref{yuk}) are generated upon integrating out the messengers. This choice can be conveniently carried out by drawing tree-level Feynman diagrams, see Fig.~\ref{Fchain}, in which a given Yukawa entry $y_{ij}^U$ is induced by the interactions in Eq.~(\ref{flavonint}) with the number of flavon insertions given by $n_{ij}^U$ in Eq.~(\ref{yuk}). The diagram dictates the required messenger couplings and therefore their $G_F$ quantum numbers. In the following, we are going to refer to these diagrams as ``chain'' diagrams, for which we introduce the  
shorthand notation:

\begin{equation}
  q_i \!-\! Q_1\! - \!\ldots \!-\!Q_{n_Q} \!-\! U_{n_U} \!-\! \ldots \!-\! U_1 \!-\! u^c_j \,.
\end{equation}
Notice that for generating a Yukawa entry with a given number of flavon insertions one can write several chain diagrams using different messengers. This ambiguity arises from the possible permutations of the Higgs and flavon insertions in Fig.~\ref{Fchain}. For instance, one can choose the position of the Higgs insertion, which corresponds to the number of ``left-handed" ($Q$) and ``right-handed" ($U,D$) messengers 
one wants to use. In particular it is possible to use only left-handed or right-handed messengers. 

All entries of the light Yukawas can then be generated using the chain diagrams. To be economic one can use the same messengers for different chains, but one has to pay attention that the resulting Yukawa matrix has full rank, which might not be the case because of possible correlations of different entries. An elegant and simple way to find the minimal number (meaning minimal number of SM representations) of messengers needed in each sector was pointed out 
in Ref.~\cite{MMM1}. In the full theory the role of the messenger couplings in Eq.~(\ref{flavonint}) is to generate three light eigenvalues 
of the full mass matrix that involves the chiral fields and the vectorlike messengers. Since the heavy eigenvalues are $\ord{M}$, the determinant
of the full mass matrix must be equal to the determinant of the light mass matrix, up to powers of $M$:
\begin{equation}
 \det M_{\rm full}^{u,d} \propto \det m_{\rm light}^{u,d} \propto \prod_I\phi_I^{n_I} v_{u,d}^3.
\end{equation}
Because every entry of the full mass matrix involves at most one power of $\phi_I$, one needs at least $N_{min}$ messengers with 
\begin{align}
\label{Nmin}
N_{min} = \sum_I n_I.
\end{align}
That is, the minimum number of messengers can be found simply by counting the powers of $\epsilon_I$ in the determinant of the Yukawa matrix.

Although this argument gives the minimal number of messengers, it is unclear which messengers have to be included. We therefore outline a simple, model-independent, procedure to derive minimal sets of fermionic messengers for each sector. One first identifies the three entries of the effective Yukawa matrix that gives the leading order contribution to the determinant. If the leading order determinant is the sum of several terms one can choose any of the summands. Then one constructs the chain diagrams for the three chosen entries using different messengers for different entries, even if they have the same quantum numbers. In this way one adds a total number of messengers that is precisely $N_{min}$. Usually there are several solutions obtained from permuting the Higgs and flavon insertions in each diagram. By counting the total number of these permutations for each of the three entries one can easily find the total number of possible UV completions with minimal number of fermionic messengers. In Appendix A we explicitly construct the Fermion messenger sector for three examples in order to illustrate the general procedure. 

The above method guarantees that the resulting Yukawa matrices are full-rank, i.e.~all fermion masses 
are generated. In general it does not ensure that other Yukawa entries besides the three chosen, and therefore the correct mixings, are generated as well. However, we checked that this is the case for the three example models of appendix A. 

Finally we want to elucidate the necessity of using different messengers for different entries. Let us assume that two Yukawa entries $y_{mn}^U$, $y_{rs}^U$ ($m \ne r, n \ne s$), contribute at leading order to the determinant and arise from chains that share a messenger. One can then integrate out all messengers but this one. The superpotential of Eq.~(\ref{flavonint}) reduces to the following form:
\begin{equation}
\label{eff-lagrangian}
W_{\rm eff} \supset M_Q \Qb Q + \alpha_i \phi \, \Qb  q_i +   \beta_j h_u  Q u^c_j \qquad  \qquad  {\rm for} \quad i= m,r \qquad j=n,s,       
\end{equation}
where $\alpha_{m,r}$ and $\beta_{n,s}$ are some effective couplings involving appropriate powers of $\epsilon_I$. By integrating out the last messenger $Q+\Qb$ one clearly obtains a rank 1 matrix (see the diagram below). 
\begin{figure}[H]
\begin{center}
\includegraphics[scale=0.8]{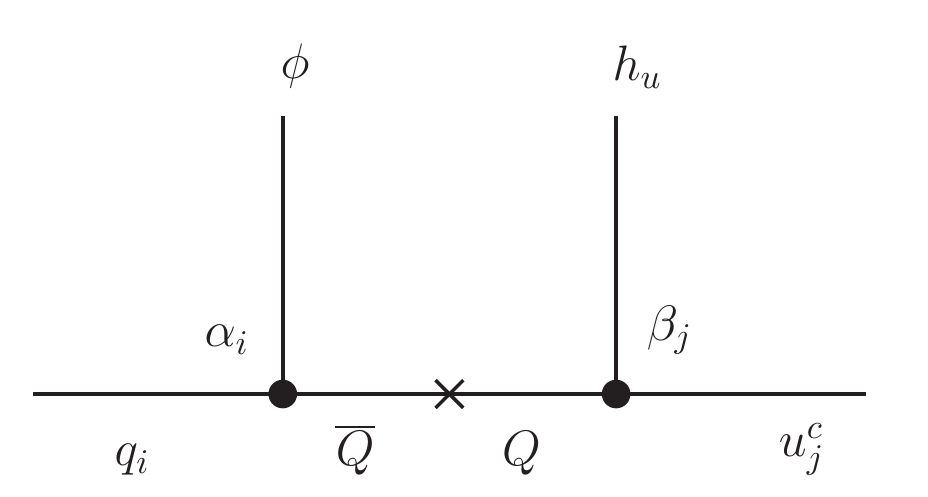}
\end{center}
\end{figure}
\noindent One can avoid this
by introducing a copy of $Q+\Qb$, i.e.~a new messenger $Q^\prime+\Qb^\prime$ with the same quantum numbers as $Q+\Qb$. This gives rise to a new contribution to the effective superpotential 
\begin{equation}
\label{eff-lagrangian2}
\Delta W_{\rm eff} \supset M_{Q^\prime} \Qb^\prime  Q^\prime +  \alpha_i^\prime \phi \, \Qb^\prime q_i +   \beta_j^\prime h_u  Q^\prime  u^c_j \qquad   {\rm for} \quad i= m,r \qquad j=n,s,              
\end{equation}
which ensures that the effective Yukawa matrix for $(q_m,q_r)$ and $(u_n^c,u^c_s)$ has rank 2. 

\subsection{UV completion with heavy scalars}
\label{HUVC}
We now discuss the UV completion of flavour models with heavy Higgs fields. In general one has to introduce certain numbers of vector-like superfields 
\begin{equation*}
H_\alpha + \overline{H}_\alpha \qquad \quad S_\alpha +\overline{S}_\alpha
\end{equation*}
where $H$ has the SM quantum number of $h_u$ and $S$ is a gauge singlet. The superpotential interactions are of the schematic form 
\begin{align}
W & \supset M_H \Hcb_\alpha H_\alpha + M_S \Sb_\alpha S_\alpha + \phi_I \left( \Hcb_\beta H_\alpha + \Sb_\beta S_\alpha + \Sb_\alpha \phi_J  \right) \nonumber \\
& + \Hcb_\alpha S_\beta h_u + \Hcb_\alpha \phi_I h_u +q_i u_j^c H_\alpha 
\end{align}
Again we dropped all couplings that are assumed to be $\ord{1}$ and take $M_H \sim M_S \sim M$. The required couplings and therefore the transformation properties of the messengers under $G_F$ can be inferred from chain diagrams like in Fig.~\ref{Hchain}. Note that for generating all Yukawa entries, in general corresponding to different charges, one needs different\footnote{In non-abelian models these fields can be part of the same multiplet.} Higgs fields coupling to $q_i u^c_j$ for each $i,j$. For generating a Yukawa entry in general one can write several diagram with different messengers, corresponding to the possible permutations of Higgs and flavon insertion in Fig.~\ref{Hchain}. In particular, it is possible to use only $H$ messengers and avoid gauge singlets. Explicit examples with HUVC can be found in Appendix A.

In the fundamental theory, small fermion masses arise from small vevs of the $H$ messengers. These vevs can be calculated by setting the messenger F-terms to zero and using the MSSM Higgs vev $\langle h_u \rangle$ as a background value. Solving the F-term equations  
\begin{align}
\partial W/\partial H_\alpha = \partial W/\partial \Hcb_\alpha = \partial W/\partial S_\alpha &= \partial W/\partial \Sb_\alpha  = 0\,, 
\end{align}
is then equivalent to integrating out the messengers by their SUSY equations of motion. This is analogous to the supersymmetric type-II see-saw \cite{typeII}, where integrating out the heavy triplet or computing the vev of its scalar component are two equivalent ways to compute the effective neutrino masses. Notice that the interpretation of small fermion masses in the fundamental theory is much easier for the Higgs UVC, since the calculation is formally the same as in the effective theory, while in the fermion messenger case one would have to diagonalise huge mass matrices.

This feature of HUVCs allows to enforce texture zeros in the Yukawa matrices in a very simple way. Although a certain Yukawa entry would be allowed by the flavour symmetry, one can set it to zero because of particular dynamics in the messenger sector. From Fig.~\ref{Hchain} it is clear that a specific Yukawa entry can only arise if the corresponding coupling to a heavy Higgs is present. If such a Higgs with the correct transformation properties under $G_F$ is missing, the entry vanishes in the fundamental theory and remains zero in the low-energy effective theory. This elegant possibility to produce texture zeros has been already outlined in~\cite{quilt}.
In Appendix A we illustrate this for the case of a U(1) flavour model. 


\section{Constraints from Perturbativity and Unification}

As we have seen, the strongly hierarchical pattern of fermion masses requires a large number of messengers with SM quantum numbers. These fields have a strong impact on the running of the SM gauge couplings, so that for a given messenger sector the messenger mass scale cannot be too far below the cutoff scale in order to avoid Landau poles. 
If the messengers are coloured (as in the case of FUVC), the strongest constraint typically arises from the running of $\alpha_3$ which is sizable already at $M_Z$. In the case of HUVC, instead the bound is set by $\alpha_2$ and is usually weaker than in the previous case. Moreover some heavy Higgs fields can be replaced by heavy singlets with no impact on running of the gauge couplings. Therefore one can expect that models with HUVC will be less constrained than FUVC models regarding the bounds from perturbativity of gauge couplings. Similarly, the messenger Lagrangian contains many couplings which are $\ord{1}$ and one has to ensure that they do not blow up in the running between the messenger scale and the cutoff. Further constraints on the messenger sector finally arise when one requires that the approximate unification of gauge couplings in the MSSM is not spoiled. This is easily ensured if the messenger fields from complete SU(5) multiplets. We are now going to discuss these issues in more detail. 

\subsection{Perturbativity of the gauge couplings}
\begin{figure}[t]
\begin{center}
\includegraphics[width=0.9\textwidth]{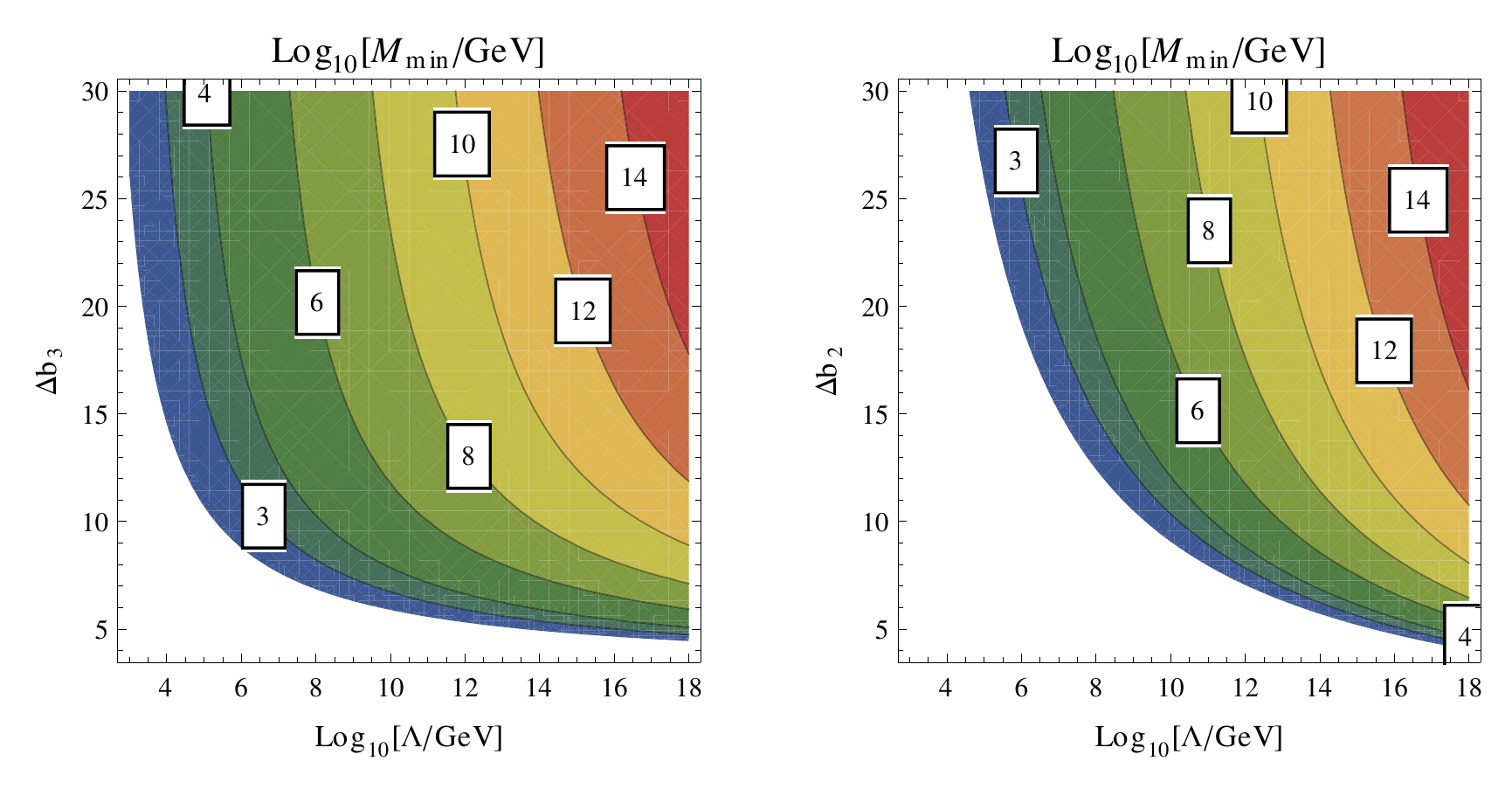}
\caption{\label{pert} Contour plot of the lower bound on the messenger scale $M_{min}$ for given $\Delta b_3$ (left), $\Delta b_2$ (right) 
requiring perturbativity up to the cutoff scale $\Lambda$ (obtained using Eq.~(\ref{pert-bound})).}
\end{center}
\end{figure}
We want to calculate the constraints on the messenger scale for a given UV completion with $N_3$ vector-like SU(3)$_c$ triplet messengers and $N_2$ vectorlike SU(2) doublet messengers living at a scale $M$. Above $M$ the 1-loop $\beta$-function coefficients $b_i$ of the gauge couplings $\alpha_i$ get modified according to 
\begin{equation}
b_i = b^{0}_i + \Delta b_i = b^{0}_i + N_i,
\end{equation}
where $b^0_i$ are the MSSM coefficients $(b^{0}_3, b^{0}_2) = (-3,1)$. Requiring that $\alpha_i$ remains in the perturbative regime ($\alpha_i \lesssim 4 \pi$) up to the cutoff scale $\Lambda$ provides a lower bound on the messenger scale $M$ given by 
\begin{align}
M \gtrsim M_{min} = \Lambda \exp{\left[ - \frac{2 \pi}{\Delta b_i} \left(\frac{1}{\alpha_i(M_Z)} - \frac{b^{0}_i}{2 \pi} \log{\frac{\Lambda}{M_Z}} \right) \right]},
\label{pert-bound}
\end{align}
where we considered 1-loop running and neglected the SUSY thresholds. Fig.~\ref{pert} shows the contours of the lower bound on the messenger scale $M_{min}$ for $\Delta b_3$ (left), $\Delta b_2$ (right) and the cutoff $\Lambda$.
Comparing the two panels, one can see that the bound is indeed stronger if the messengers are coloured.

In the case of a FUVC, one can get, for a given effective model, the minimal number of triplets directly from Eq.~(\ref{Nmin}). Taking into account both up- and down sector one has $N_3 = N_{min}^u + N_{min}^d$, provided that only right-handed messengers are used in both sectors. If also left-handed messengers are used, in general the number of triplets is larger, unless both isospin components of the quark doublet serve as messengers. In turn, one can use the perturbativity bound as a criterion for defining the ``minimal'' fermionic messenger sector as the solution with the least number of color triplets, i.e.~the messenger sector least constrained by perturbativity. This notion can reduce drastically the degeneracy of minimal messenger sectors since only few solutions efficiently unify up and down sectors. 

The UV completions for the example models in the Appendix are chosen to fulfil this criterion of minimality, and can therefore be used to illustrate the minimal bounds on the messenger scale for typical flavour models. We summarise the results in Table~\ref{compFH}.

\begin{table}[h]
\centering
\begin{tabular}{|c||c|c|}
Model & FUVC & HUVC \\
& $\Delta b_3$  \hspace{0.5cm} $M_{min}$ &  $\Delta b_2$  \hspace{0.5cm} $M_{min}$\\
\hline
U(1) & 19 \hspace{0.7cm} $10^{14}$ & 11 \hspace{0.7cm} $10^{12}$ \\
U(1) $\times$ U(1)$'$ & 8 \hspace{0.7cm} $10^9$ & 11 \hspace{0.7cm} $10^{12}$ \\
SU(3) & 12 \hspace{0.7cm} $10^{12}$ & 6 \hspace{0.75cm} $10^{7}$  \\
\hline
\hline
\end{tabular}
\caption{\label{compFH}Comparison of HUVC and FUVC for the three example models in Appendix A. $M_{min}$ denotes the lower bound on $M$ in GeV requiring perturbativity up to $M_{\rm Planck}$. }
\end{table}
\noindent It is interesting to notice that, despite the large number of additional fields charged under the SM gauge group, a theory perturbative up to the Planck scale is still achievable with a messenger sector living far below the GUT scale.

\subsection{Perturbativity of the $\ord{1}$ couplings}

The couplings in the messenger sector are by construction $\sim \ord{1}$, but they cannot be too large if the theory should remain perturbative up to the cutoff $\Lambda$.
One can easily estimate the corresponding bounds considering the typical RGE of a generic superpotential coupling $\lambda$.
Let us consider for simplicity only the unavoidable contribution to the $\beta$-function that is proportional to the third power of the coupling itself:\footnote{In general there will be other terms proportional to different Yukawas, of the kind $\lambda \lambda^{\prime\,2}$, that would only make the bound stronger, and gauge terms, $g_i^2\lambda$, which are negligible in the large coupling regime we are interested in, $\lambda \gtrsim 1$.}
\begin{align}
 (4 \pi)^2 \frac{d}{dt} \lambda = C \lambda^3\,.
\end{align}
Here the coefficient $C$ is given by group theoretical factors that depend on the quantum numbers of the fields appearing in the coupling. For some typical couplings that appear in the chains of Figs.~\ref{Fchain} and \ref{Hchain} one finds:
\begin{align}
 \lambda\, Q_\alpha \overline{Q}_\beta \phi ~ & \Rightarrow ~ C=8 \nonumber \\
\lambda\, U_\alpha \overline{U}_\beta \phi ~ & \Rightarrow ~ C=5 \nonumber \\
\lambda\, H_u^I H_d^J \phi ~ & \Rightarrow ~ C=4 \nonumber.
\end{align}
Requiring the absence of a Landau pole below the cutoff (i.e.~$1/\lambda(\Lambda) >0$) 
gives an upper bound on the value of the coupling at the messenger scale $M$:
\begin{align}
 \lambda (M) \lesssim \lambda_{max} =\left(\frac{C}{ 8\pi^2} \log\frac{\Lambda}{M}\right)^{-1/2}\,.
\end{align}
One can get very conservative bounds considering a short running with $\Lambda/M = 10$. In this case we find
$\lambda_{max} = (2.9,~2.6,~2.1)$ for $C = (4,~5,~8)$.
\subsection{Unification}
One might want to require that the apparent unification of the MSSM gauge couplings is not spoiled by the presence of the messenger sector living at an intermediate scale. This in general gives rise to additional constraints on the UV completion. As is well known, MSSM unification is exactly preserved at 1-loop if the additional fields 
form complete multiplets of SU(5), independent of the scale where they live.\footnote{For sets of fields that do not form complete SU(5) multiplets but
still maintain MSSM unification, see Ref.~\cite{Magic}.} 
Indeed the embedding of the messengers in 
SU(5) multiplets is straightforward provided that the flavour symmetry commutes with SU(5), i.e.~the effective theory can be written in a SU(5) invariant way.

For FUVC one has now to write the chain diagrams with messengers transforming as $\bf 10 + \overline{10}$ and $\bf \overline{5} + 5 $ of SU(5). This leads to a further reduction of the degeneracy of messenger sectors that minimise perturbativity constraints. Indeed the UV completions for the SU(5) compatible examples in the Appendix are the unique solutions satisfying this strong requirement.  

For HUVC the Higgs messengers can be simply embedded in $\five+\bar{\five}$ representations. In this case also the perturbativity bounds on $\alpha_3$ become relevant. Notice that the introduction of new Higgs color triplets at the scale $M$ potentially gives rise to additional contributions to proton decay. While in general this might lead to new constraints on $M$, for the SU(5) compatible models in the appendix this is not the case. The reason is that in these models any tree-level diagram which generates the dangerous dimension 5 proton decay operators ($qqq\ell$ and $uude$) via a chain of heavy Higgs triplets has to pass through the ordinary coloured triplets associated with the light Higgs doublets. Therefore proton decay is suppressed by the mass of these triplets together with small Yukawas precisely as in ordinary SUSY GUTs. This means that the new triplets can be light as long as the Higgs coloured triplets live at the GUT scale.


\section{Constraints from SUSY-induced flavour violation}
\label{pheno}

In the previous section we have shown that a flavour theory that remains perturbative up to the Planck scale requires very heavy messengers, in the order of $M = 10^{10}$ GeV. This implies that all direct effects at low energy can be neglected, since they are suppressed by powers of $M$. However, in the MSSM even such high scales can have an important impact on TeV scale physics due to the presence of light SUSY particles. Their soft masses are determined by the underlying mechanism of SUSY breaking, and are usually generated at very high scales as well. This means that the messenger sector can interfere with the SUSY breaking sector. In particular the messenger sector strongly violates flavour universality by construction, and therefore can easily induce flavour violation in the sfermion masses with drastic consequences for low-energy observables~\cite{Buras-Paride}.

\subsection{Tree-level Effects}
A common approach in the literature is to assume high-scale  SUSY breaking like Gravity Mediation and apply a spurion analysis to determine the structure of the sfermion mass matrices below the messenger scale. If the messenger scale $M$ is much below the SUSY breaking scale $M_S$, all flavour-violating effects in the soft terms arise dominantly from the messenger sector (see e.g. \cite{Kadota:2010cz}). This is because at the SUSY breaking scale flavour-violating effects are suppressed by powers of $\phi/ M_S$ instead of $\phi/ M$ and therefore are negligible. All soft mass terms for chiral fields and messengers are given in the flavour symmetric limit and read schematically 
\begin{align}
K \supset \frac{X^\dagger X}{M_S^2} \left( a_i q^\dagger_i q_i + b_\alpha  Q^\dagger_\alpha Q_\alpha + c_{i \alpha} q_i^\dagger Q_\alpha + d_\beta H^\dagger_\beta H_\beta \hdots  \right) + \ord{\phi/M_S},
\end{align}  
where we canonically normalised the fields,  $X$ denotes the SUSY breaking spurion and $Q, H$ denote Fermion and Higgs messengers, respectively. 

In general integrating out the messenger fields generates the tree-level flavour structure of the soft masses as expected by the flavour symmetry (controlled by powers of $\phi/M$). However this effective structure also depends on the details of the messenger sector and the SUSY breaking mechanism. For example it is clear that for Higgs messengers no off-diagonal sfermion mass terms can be generated in the flavour basis. Instead Fermion messengers do mix with light fields and this mixing typically generates off-diagonal sfermion masses. This is however not the case in Gauge Mediation where sfermion masses are universal among fields with same gauge quantum numbers ($a_i = b_\alpha, c_{i \alpha} = 0 $), and therefore stay universal in every basis. 
Still, even in such cases flavour universality is broken by radiative effects as we are going to discuss in the following. 

\subsection{Radiative Effects}
Let us assume a mechanism of SUSY breaking that generates universal sfermion masses at the scale $M_S$. This can be easily realised if the underlying dynamics is flavour-blind, for instance in Gauge Mediation. If $M_S$ is below the flavour messenger scale $M$, the RG running down to low energies approximately preserves universality, because it is broken only by the Yukawa interactions that are small for the relevant transitions between 1st and 2nd generation. If instead the SUSY breaking scale is above the messenger scale, universality is spoiled by messenger loop corrections\footnote{This effect has been discussed for a model with accidental flavour symmetries and Fermion messengers~\cite{Conseq}.}, since the interactions of sfermions and messengers strongly violate $SU(3)^5_F$. While in abelian flavour models this breaking is large, being through $\ord{1}$ couplings in the Lagrangian, in non-abelian models it is due to small flavon vevs. Still, in simple non-abelian models \cite{models-u2,models-su3,Ross} the 1-2 sector universality breaking can be sizable (of the order of the Cabibbo angle squared) and in 1-3 and 2-3 sector 
even of $\ord{1}$ . 

The RG effects from the messenger sector destroy universality of sfermion masses in two ways, by directly generating an off-diagonal entry and by splitting degenerate masses on the diagonal that is then converted to an off-diagonal entry in the mass basis.\footnote{This is perfectly analogous to the RG induced flavour violation in SUSY seesaw~\cite{Borzumati:1986qx} and SU(5) models~\cite{Barbieri-Hall}.}As we will see, the second contribution is always larger or equal than the first one for large classes of flavour models. For such models one can therefore obtain constraints on the light fermion rotations depending on the size of the diagonal splitting. 

In this section we estimate these constraints for the case of abelian and simple non-abelian flavour symmetries. The starting point is a universal sfermion mass matrix at the SUSY scale $M_S$ where for simplicity we restrict to 1st and 2nd generation
\begin{align}
\tilde{m}_{ij}^2 (M_S)= \begin{pmatrix} \tilde{m}_0^2 & 0 \\ 0 &  \tilde{m}_0^2 \end{pmatrix}.
\end{align}
When this mass matrix is evolved down to the scale $M$ where the messengers decouple, all entries receive corrections $\Delta \tilde{m}^2_{ij}$ which are determined by the RG equations
\begin{align}
\tilde{m}_{ij}^2 (M)= \begin{pmatrix} \tilde{m}_0^2 + \Delta \tilde{m}^2_{11}& \Delta \tilde{m}^2_{12} \\ \Delta \tilde{m}^2_{21} &  \tilde{m}_0^2 + \Delta \tilde{m}^2_{22} \end{pmatrix}.
\end{align}
The final evolution to the soft SUSY breaking scale scale is determined by gauge (thus flavour universal) terms and by 1st and 2nd family MSSM Yukawa couplings that can be neglected: 
\begin{align}
\tilde{m}_{ij}^2 (\tilde{m})\approx \tilde{m}_{ij}^2 (M) + \delta_{ij}~C_a M_a^2 (M),
\label{soft}
\end{align}
where $M_a$ are gaugino soft masses.
The 1-2 entry in the super-CKM basis is then approximately given by 
\begin{align}
\label{2contrib}
\tilde{m}_{12}^2 \approx \Delta \tilde{m}^2_{12} + \left( \Delta \tilde{m}^2_{22} - \Delta \tilde{m}^2_{11} \right) \theta_{12},
\end{align}
where $\theta_{12}$ denotes the (complex) rotation angle in the sfermion sector under consideration. For simple flavour models like $U(1), U(1)^2$ or $SU(3)$ it is easy to see the second term is always larger or equal than the first one, provided the rotation angle does not vanish. For example in a $U(1)$ model one has for left-handed down squarks 
\begin{align}
\label{m12rel}
\Delta \tilde{m}_{12}^2 \propto \eps^{ q_1 - q_2} \tilde{m}^2, \quad  \Delta \tilde{m}^2_{22} - \Delta \tilde{m}^2_{11} \propto \tilde{m}^2, \quad \theta_{12} \sim \frac{y^D_{12}}{y^D_{22}} \propto \frac{\eps^{q_1 + d_2}}{ \eps^{q_2 + d_2}} =  \eps^{q_1 - q_2} ,
\end{align}
 where $q_i, d_i$ denote the U(1) charge of the corresponding superfields. A non-abelian example will be discussed below. As we shall see later $\tilde{m}^2 \approx \tilde{m}^2_0 $, that is the loop corrections are of the same order as the tree-level effects.  Therefore in the following we restrict our attention on the second term in Eq.~(\ref{2contrib}). We neglect the possibility that the two terms cancel since in general they involve different $\ord{1}$ coefficients. In appendix B we illustrate this issue in an explicit example. 

The diagonal splitting $  \Delta \tilde{m}^2_{22} - \Delta \tilde{m}^2_{11}$ depends on the RG running that can be estimated in the leading-log approximation. It depends on all interactions of sfermions and messengers. Since these couplings are $\ord{1}$ the RG coefficients are in general large.  As a conservative estimate we take for the total RG contribution a factor 10 into account (besides the loop factor and the logarithm), a typical value one finds in concrete models (see appendix \ref{RGE-appendix}). In abelian models there is no extra suppression, because 1st and 2nd generation sfermions couple with different $\ord{1}$ couplings to the messengers. Instead in non-abelian models 1st and 2nd family sfermions can be embedded in the same representation under the flavour group, which implies that their couplings to messengers are universal, except for small symmetry breaking effects which lead to additional suppression. Such breaking is model dependent and can be even of the order $\eps^2$ in the 1-2 sector. To estimate the additional suppression  factor we consider the case of a generic SU(3) model with all quarks transforming as a ${\bf 3}$ and flavons as ${\bf \overline{3}}$. The flavons get hierarchical vevs that induce the quark masses. The flavon vev responsible for universality breaking in the $i$-$j$ sector is therefore roughly given by the square root of the Yukawa coupling $\sqrt{y_{jj}}$. The situation is similar in the case of some U(2)$_F$ flavour models \cite{models-u2}, where transitions in 1-3 and 2-3 sector are always unsuppressed.  We find that the additional suppression factor for the sfermion mass splitting $\Delta_{ij} \equiv \tilde{m}^2_{ii} -  \tilde{m}^2_{jj} $ is given by $y_{jj}$:
\begin{table}[H]
\centering
\begin{tabular}{|c || c | c|}
Mass splitting & Suppression factor in SU(3)$_F$ & Suppression factor in U(2)$_F$\\
\hline
\hline
$\Delta_{13}^U $ & $\ord{1}$ & $\ord{1}$ \\  
$\Delta_{23}^U $& $\ord{1}$ & $\ord{1}$ \\  
$\Delta_{12}^U $ & $\eps^4$  & $\eps^4$ \\  
\hline
$\Delta_{13}^D $& $ \eps^3 \tan{\beta}$ & $\ord{1}$\\  
$\Delta_{23}^D $&  $ \eps^3 \tan{\beta}$ & $\ord{1}$\\  
$\Delta_{12}^D $ &  $ \eps^5 \tan{\beta}$ &  $\eps^5 \tan{\beta}$ \\  
\hline
\hline
\end{tabular}
\caption{\label{suppNA}Additional suppression factors of diagonal sfermion mass splittings $\Delta_{ij} \equiv \tilde{m}^2_{ii} -  \tilde{m}^2_{jj} $ in simple non-abelian models. }
\end{table}
Let us finally illustrate that also in the non-abelian case the second term in Eq.~(\ref{2contrib}) gives a good estimate for the total contribution. 
Since we assume left-handed and right-handed superfields to form triplets of the non-abelian symmetry, the generated $\Delta \tilde{m}_{12}^2$ and 
$\Delta \tilde{m}^2_{22} - \Delta \tilde{m}^2_{11}$ are proportional to two flavon insertions, like the Yukawa entries.
For left-handed down squarks, we then have:
\begin{align}
\Delta \tilde{m}_{12}^2 \lesssim y^D_{12} \,\tilde{m}^2, \quad  \Delta \tilde{m}^2_{22} - \Delta \tilde{m}^2_{11} \sim y^D_{22}\, \tilde{m}^2, \quad \theta_{12} \sim \frac{y^D_{12}}{y^D_{22}},
\label{m12relNA}
\end{align}
where the inequality in the first expression accounts for the fact that additional symmetries typically lead to a further suppression of soft masses with respect to the naively expected size $\propto y^D_{12}$. Note that such a suppression can occur only for off-diagonal sfermion masses. From Eq.~(\ref{m12relNA}), we then see that also in this case $(\Delta \tilde{m}^2_{22} - \Delta \tilde{m}^2_{11}) \theta_{12} \gtrsim \Delta \tilde{m}^2_{12}$.

In summary, for the estimation of the mass splitting we consider the case of abelian flavour symmetries, and keep in mind that in non-abelian models one can have additional suppression. Still the abelian case can be relevant in non-abelian models for sfermions transforming as singlets under $G_F$, but possibly charged under additional U(1) factors (e.g.~\cite{models2,Antusch:2010es}). 
We can then estimate the off-diagonal sfermion mass at leading log by:
\begin{align}
\label{m12est}
\tilde{m}^2_{12} \approx \left(  \Delta \tilde{m}^2_{22} - \Delta \tilde{m}^2_{11} \right) \theta_{12} \approx   \theta_{12}  \frac{\tilde{m}_0^2}{16 \pi^2} 10 \log\frac{M_S}{M} .
\end{align}
Note that this estimate is roughly of the same order as one would expect for a tree-level sfermion mass matrix only constrained by the flavour symmetry. If the rotation angle does not vanish, it gives the leading contribution to the off-diagonal entry in the super-CKM basis. It only depends on the rotation angles, whereas the off-diagonal entry in the flavour basis depends directly on the specific flavour symmetry~\cite{LPR} and can be affected by the messenger sector. 

The corresponding mass insertion $\delta_{12} \equiv \tilde{m}^2_{12}/\sqrt{\tilde{m}^2_{11}\tilde{m}^2_{22}}$ is then given by:
\begin{align}
\label{predMI}
\delta_{12}^{\rm ab.} & \approx  \frac{ \theta_{12}}{16 \pi^2} 10~ \mathcal{R}\log\frac{M_S}{M}\,,    
\end{align}
where the factor $\mathcal{R}$ parameterizes the possible suppression due to the gaugino-driven running of the diagonal entries, cf.~Eq.~(\ref{soft}). 
$\mathcal{R}$ is typically $\ord{1}$ in the case of sleptons, while for squarks it ranges from $\ord{1}$ down to $\ord{0.1}$ in the case of large high-energy
gluino mass, $M_3 \gg \tilde{m}_0$.

Similarly, radiative effects given by messenger loops will induce flavour-violating entries in the A-term matrices in the super-CKM basis, even if they vanish at high energy or are aligned to the corresponding Yukawas. Given that the flavour structures of the A-terms and the Yukawas are the same, while their $\beta$-function coefficients differ by $\ord{1}$ factors, 
the radiatively generated LR mass-insertions, $(\delta^f_{LR})_{ij} \equiv (A_f m_f)_{ij}/\tilde{m}^2$ can be estimated to be:
\begin{align}
\label{predMI-LR}
(\delta^f_{LR})_{ij} &  \approx \frac{1}{16 \pi^2} a_{ij} Y^f_{ij} ~\frac{v}{\tilde{m}} \log\frac{M_S}{M},
\end{align}
where $a_{ij}$ are $\ord{1}$ coefficients, $v$ is the EWSB vev and $\tilde{m}$ the low-energy squark/slepton mass.
\begin{table}[t]
\centering
\setlength{\extrarowheight}{1.8pt}
\begin{tabular}{|c ||c c |}
\hline
\hline
$(\delta^D_{XX})_{12}$ &  $9.2\times 10^{-2}$ [Re] &   $1.2\times 10^{-2}$ [Im]   \\  
$\langle\delta^D_{12}\rangle$ &  $1.9\times 10^{-3}$ [Re]  &   $2.6\times 10^{-4}$ [Im]   \\
$(\delta^D_{LR})_{12}$ &  $5.6\times 10^{-3}$ [Re] &   $7.4\times 10^{-4}$ [Im]   \\  
\hline
$(\delta^U_{XX})_{12}$   &  $1.0 \times 10^{-1}$ [Re]   &   $6.0 \times 10^{-2}$ [Im]    \\
 $\langle\delta^U_{12}\rangle$  & $6.2\times 10^{-3}$ [Re]    &  $4.0 \times 10^{-3}$ [Im]    \\
$(\delta^U_{LR})_{12}$ &  $1.6\times 10^{-2}$ [Re] &   $1.6\times 10^{-2}$ [Im]   \\  
\hline
$(\delta^D_{XX})_{13}$ &  $2.8\times 10^{-1}$ [Re]  &   $6.0\times 10^{-1}$ [Im]   \\
$\langle\delta^D_{13}\rangle$ & $4.2\times 10^{-2}$ [Re]   &   $1.8\times 10^{-2}$ [Im]  \\
$(\delta^D_{LR})_{13}$ &  $6.6\times 10^{-2}$ [Re] &   $1.5\times 10^{-1}$ [Im]   \\  
\hline
$(\delta^E_{LL})_{12}$      &  \multicolumn{2}{c|}{$2.8\times 10^{-3}$~~ $[5.7\times 10^{-4}]$}    \\
$(\delta^E_{RR})_{12}$ &  \multicolumn{2}{c|}{$2.3\times 10^{-2}$ ~~$[4.6\times 10^{-3}]$}  \\  
$\langle\delta^E_{12}\rangle$  &   \multicolumn{2}{c|}{$1.8\times 10^{-3}$ ~~$[3.8\times 10^{-4}]$}   \\
$(\delta^E_{LR})_{12}$ &    \multicolumn{2}{c|}{$1.7\times 10^{-5}$~~ $[3.4\times 10^{-6}]$}   \\  
\hline
\hline
\end{tabular}
\caption{\label{MIbounds} Bounds on flavour-violating mass-insertions. 
Here $\langle\delta^f_{ij}\rangle \equiv \sqrt{(\delta^f_{LL})_{ij}  (\delta^f_{RR})_{ij}}$ and $X=L,R$.
Values in [ ] denote expected future bounds. See the text for details.}
\end{table}
\subsection{Numerical Discussion}
The expressions above can be compared to the various bounds on the mass insertions obtained from FCNC and LFV processes. In Table~\ref{MIbounds} we collected the quark sector constraints from the existing literature \cite{Masiero-Vives-Vempati,Ciuchini:2007cw,GinoNir}, for the following
reference values of squark and gluino masses: $m_{\tilde q} =1$ TeV, $m_{\tilde g}/m_{\tilde q} =1$.
For the leptonic sector we used the new exclusion limit on $\mu \to e \gamma$ from the MEG collaboration~\cite{MEG} to update the existing bounds on $(\delta^E_{LL})_{12}$, $(\delta^E_{RR})_{12}$ and $(\delta^E_{LR})_{12}$. We performed a random variation of the SUSY parameters in the following ranges: $m_{\tilde \ell} = [100,1000]$ GeV, 
$M_1 = [50,500]$ GeV, $M_2 = [100,1000]$ GeV, $\mu = [100,2000]$ GeV, $\tan\beta = [5,15]$; then, we have taken as bound for a given mass-insertion the value
for which 90\% of the points of the scan are excluded by BR($\mu \to e \gamma$) (which has been computed using the expressions in \cite{Paride}). 
Doing like that, we neglected cancellations among different contributions larger than roughly 10\%.
\begin{table}[t]

\setlength{\extrarowheight}{3pt}
\centering
\begin{tabular}{|c||c|c|c|c|c|}
rotation angle & $M_S/M = 10^{8} $ & $M_S/M = 10 $\\
\hline
\hline

$\theta^{DL}_{12},\theta^{DR}_{12} $&  $7.9\times 10^{-2}~{\rm [Re]}$~~~ $1.0\times 10^{-2}~{\rm [Im]}$& $6.3\times 10^{-1}~{\rm [Re]}$~~~ $8.2\times 10^{-2}~{\rm [Im]}$\\

$\langle\theta^D_{12} \rangle$&  $1.6 \times 10^{-3}~{\rm [Re]}$~~~ $2.2 \times 10^{-4}~{\rm [Im]}$&  $1.3 \times 10^{-2}~{\rm [Re]}$~~~ $1.8 \times 10^{-3}~{\rm [Im]}$\\

\hline

$\theta^{UL}_{12},\theta^{UR}_{12} $&  $8.6\times 10^{-2}~{\rm [Re]}$~~~ $5.1 \times 10^{-2}~{\rm [Im]}$& $6.9\times 10^{-1}~{\rm [Re]}$~~~ $4.1\times 10^{-1}~{\rm [Im]}$\\

$\langle \theta^{U}_{12}\rangle   $&  $5.3 \times 10^{-3}~{\rm [Re]}$~~~ $3.4\times 10^{-3}~{\rm [Im]}$& $4.3 \times 10^{-2}~{\rm [Re]}$~~~ $2.7\times 10^{-2}~{\rm [Im]}$\\

\hline

$\theta^{DL}_{13},\theta^{DR}_{13} $&  $2.4\times 10^{-1}~{\rm [Re]}$~~~ $5.1\times 10^{-1}~{\rm [Im]}$& -  \\

$  \langle\theta^{D}_{13}\rangle $&  $3.6 \times 10^{-2}~{\rm [Re]}$~~~ $1.5 \times 10^{-2}~{\rm [Im]}$& $2.9 \times 10^{-1}~{\rm [Re]}$~~~ $1.2 \times 10^{-1}~{\rm [Im]}$\\

\hline
\hline

$\theta^{EL}_{12} $&  $2.4\times 10^{-3} \quad [4.9 \times 10^{-4}]$&  $1.9\times 10^{-2} \quad [3.9 \times 10^{-3}]$\\
$\theta^{ER}_{12} $&   $2.0\times 10^{-2} \quad [3.9 \times 10^{-3}]$&  $1.6\times 10^{-1} \quad [3.2 \times 10^{-2}]$\\
$\langle\theta^{E}_{12}\rangle $&   $1.5\times 10^{-3} \quad [3.3 \times 10^{-4}]$&  $1.2\times 10^{-2} \quad [2.6 \times 10^{-3}]$\\

\hline
\hline
\end{tabular}
\caption{\label{abSUSYtable}Constraints on rotation angles in abelian flavour models, obtained using Eq.~(\ref{predMI}) with $\mathcal{R}=1$ and the bounds of Table~\ref{MIbounds} (corresponding to $m_{\tilde q}=1$ TeV). The angles without $L,R$ specification are the geometric mean of both, e.g.  $\langle\theta^D_{12}\rangle \equiv  \sqrt{ \theta^{DL}_{12} \theta^{DR}_{12} }$. For the leptonic angles $\theta^{E}_{12}$ we show in brackets also the expected future bound provided that MEG will improve the limit on BR($\mu \to e \gamma$) down to $10^{-13}$.}
\end{table}

Since the estimated effect in Eq.~(\ref{predMI}) depends only on the rotation angle and the ratio of SUSY and messenger scale, for a given ratio one obtains an upper bound on the real and imaginary part of the rotation angle. This bound depends logarithmically on the ratio of scales for which we consider the two extreme cases $M_S/M = 10$ and $M_S/M = 10^{8}$, which in Gravity Mediation corresponds to messengers at $ M \approx 10^{17}$ GeV and $M \approx 10^{10}$ GeV, the latter representing the typical minimal value satisfying perturbativity constraints. The results are summarised in Table~\ref{abSUSYtable}, for the case of an abelian flavour symmetry. Bounds for non-abelian models can be obtained taking into account the additional suppression factors provided in Table~\ref{suppNA}. 

This table can be used to estimate the constraints on the Yukawa matrix (valid up to unknown $\ord{1}$ coefficients) in a given SUSY breaking scenario. 
They are unavoidable whenever the SUSY breaking scale is above the messenger scale $M_S > M$, which includes mSUGRA. As one can see, these bounds are quite strong although they hold for pretty general flavour models. They put abelian flavour models in mSUGRA scenarios in trouble, since at least either $\theta^{UL}_{12}$ or $\theta^{DL}_{12}$ must account for the Cabibbo angle, i.e.~must be $\ord{\eps} \approx 0.2$. In realistic models the constraints are even stronger, since typically they are compatible with $SU(5)$, implying $\theta^{UL} \approx \theta^{UR}$, $\theta^{DL}  \approx \theta^{ER}$, $\theta^{DR}  \approx \theta^{EL}$. The constraints are less severe in non-abelian models (that gives at least an additional suppression $\sim \eps^2$ in the 1-2 sector), but the small breaking of universality is partially compensated by the fact that in these models fermion mass matrices are typically symmetric and therefore $\theta^L \sim \theta^R$. Still, non-abelian models are definitively preferred from what concerns the effect that we have discussed here.

As an illustration, we compare the obtained bounds with the rotation angles predicted in the U(1) and SU(3) models of Appendix A in Table~\ref{MIs}. 
\begin{table}[h]
\setlength{\extrarowheight}{3pt}

\centering
\begin{tabular}{|c||c|c|}

rotation angle & U(1) & SU(3) \\
\hline
\hline
$\theta^{DL}_{12},\theta^{DR}_{12} $ & $\eps$ & $\eps \,  (\eps^3 )$\\
$\langle \theta_{12}^{D} \rangle$ & $ \eps $ & $\eps \,  (\eps^3 )$\\
\hline
$\theta^{UL}_{12},\theta^{UR}_{12} $ & $\eps$ & $\eps^2 \, (\eps^6)$\\
$\langle \theta_{12}^{U} \rangle$ & $ \eps $ & $\eps^2 \, (\eps^6)$\\
\hline
$\theta^{DL}_{13}$ & $\eps^3$ & $\eps^3 \,  (\eps^3 )$\\
$\theta^{DR}_{13} $ & $\eps$ & $\eps^3 \,  (\eps^3 )$\\
$\langle \theta_{13}^{D} \rangle$ & $ \eps^2 $ & $\eps^3 \,  (\eps^3 )$\\
\hline
$\theta^{EL}_{12}, \theta^{ER}_{12}$ & $\eps$ & $\eps \,  (\eps^3 )$\\
$\langle \theta_{12}^{E} \rangle$ & $ \eps $ & $\eps \,  (\eps^3 )$ \\
\hline
\hline
\end{tabular}
\caption{\label{MIs}Rotation angles for example models of the Appendix A with $\eps \approx 0.2$. For the non-abelian case we included in parentheses the total effective angle using the additional suppression factor in Table~\ref{suppNA}.}
\end{table}
For the leptonic rotation angles we assumed the SU(5) relations $\theta^{EL}  \approx \theta^{DR}$, $\theta^{ER}  \approx \theta^{DL}$. Comparing to the bounds given in Table~\ref{abSUSYtable}, we see that the U(1) model (as any abelian model) is seriously challenged by the 1-2 sector 
constraints. Even considering an additional suppression $\mathcal{R}=\ord{0.1}$ from vanishing scalar masses at $M_S$ and small CPV phases, 
the bounds can be evaded only for a quite heavy SUSY spectrum, $m_{\tilde q} \simeq 2$ TeV (notice that the bounds in Table~\ref{abSUSYtable} scale like 
$m_{\tilde q}/(1~{\rm TeV})$). With this setup, the U(1) model should still exhibit sizable deviation from the SM in $K-\overline{K}$, $D-\overline{D}$ mixing, as well as
a rate for $\mu\to e\gamma$ in the reach of the MEG experiment.
On the other hand, the non-abelian example is still perfectly compatible with the bounds. Notice however that large effects in SU(3) models for $K-\bar{K}$ CP violation
and LFV are still possible, provided that the SUSY masses are not too heavy.\footnote{The phenomenological implications of some SU(3) models have been recently discussed in \cite{su3flavour}.}


\section{Conclusions}

In this paper, we have discussed the general features of the UV completions of SUSY flavour models.
We have analysed in detail the structure of the messenger sector of this kind of models,
which contains vector-like superfields that mix either with light fermions or with light Higgs fields.
In the latter case, it is particularly simple to obtain texture zeros in the Yukawa matrices,
just by removing certain messengers from the theory, without modifying the transformation properties of the SM fields.
Our discussion on the structure of the messenger sector does not rely on SUSY and can be easily generalised
to non-supersymmetric scenarios.

In general, the messenger sector contains many new fields charged under the SM gauge group. As a consequence,
requiring the theory to remain perturbative up to high-energy scales
forces the masses of the messengers to be far above the TeV scale, typically $\ord{10^{10}}$ GeV for the example models we have considered.
This implies that the messengers have no direct impact on low-energy observables.
Still, their presence affects the RG running of the sfermion masses.
We have emphasised that the radiative generation of sfermion mass-splittings and misaligned A-terms is unavoidable whenever the flavour symmetry breaking scale is lower than the SUSY breaking scale, like in Gravity Mediation. We have quantitatively estimated this RG effect and found it comparable to the tree-level off-diagonal sfermion masses as expected by the flavour symmetry. Therefore the strong flavour constraints cannot be evaded even under the assumption of universal soft terms. These constraints depend only on the diagonal mass splitting and the rotation angles, and therefore apply to large classes of flavour models. In Table~\ref{abSUSYtable}
we have provided bounds on the rotation angles of the light fermions that are valid in any abelian flavour model and can be easily extended to simple non-abelian models.
We find that abelian models are strongly constrained and hence it is difficult to marry them to Gravity Mediation with SUSY
at the TeV scale.
Even though it is well-known that abelian models induce large flavour changing effects, still we find it remarkable
that this remains true even under the strong assumption of universal soft masses at the SUSY breaking mediation scale, as a consequence of the presence of
flavour messengers affecting the RG running. Not surprisingly, in non-abelian models the sfermion mass-splittings can be suppressed with respect to the abelian case
by small flavon vevs and ease the bounds on the rotation angles. However, large effects for Kaon and LFV observables are still possible,
provided that the SUSY spectrum is not too heavy. 

\section*{Acknowledgements}
We would like to thank C.~Hagedorn, A.~Romanino and G.~G.~Ross for useful discussions. 
The Feynman diagrams have been drawn using JaxoDraw \cite{Jaxodraw}.
We thank the Theory Division of CERN for hospitality during several stages of this work. 
L.C.~and R.Z.~are grateful to the Institute of Theoretical Physics of the University of Warsaw for kind hospitality and financial support 
during their stays in Warsaw.
S.P.~and R.Z.~acknowledge support of the TUM-IAS funded by the German Excellence Initiative.
This work has been partially supported by the contract PITN-GA-2009-237920 UNILHC 
and the MNiSzW scientific research grant N202 103838 (2010-2012).

\appendix

\section{Example Models}
In this Appendix we explicitly construct the UV completions both with fermion and Higgs messengers for three example models: U(1), U(1)$\times$U(1)$^\prime$ and SU(3).
\subsection{U(1)$_{\Hc}$}
\label{sec:u1}
The first example is based on a U(1)$_{\Hc}$ flavour symmetry. 
We take for the charges of the MSSM superfields~\cite{CKLP}:
\begin{align}
 q_{1,2,3} &~:~~ (3,2,0) \nonumber \\
 u^c_{1,2,3} &~:~~ (3,2,0) \nonumber \\
 d^c_{1,2,3} &~:~~ (2,1,1), \,
\label{u1charges}
\end{align}
the Higgs fields are neutral and a single flavon is introduced with charge $\Hc(\phi)=-1$. This gives rise to the effective Yukawa matrices: 
\begin{align}
Y_u & \sim 
\begin{pmatrix}
\eps^6 & \eps^5 & \eps^3 \\
\eps^5 & \eps^4 & \eps^2 \\
\eps^3 & \eps^2 & 1 \\
\end{pmatrix}
& Y_d & \sim 
\begin{pmatrix}
\eps^5 & \eps^4 & \eps^4 \\
\eps^4 & \eps^3 & \eps^3 \\
\eps^2 & \eps & \eps \\
\end{pmatrix}.
\label{u1yuk}
\end{align}
The correct hierarchy of fermion masses and mixing can be achieved choosing the expansion parameter of the order of the Cabibbo angle, $\epsilon\sim 0.23$.

\subsubsection{Fermion UVC}

The determinants of the Yukawas matrices in Eq.~(\ref{u1yuk}) are 
\begin{align}
 \det{Y_u} & \sim \epsilon^{\sum_i \Hc(q_i) + \Hc(u_i)} = \epsilon^{10} &  \det{Y_d} & \sim \epsilon^{\sum_i \Hc(q_i) + \Hc(d_i)} = \epsilon^9,
\end{align}
where $\Hc(f_i)$ is the charge of the fermion $f_i$. 
According to Eq.~(\ref{Nmin}), one needs in total $\sum_i \Hc(q_i) + \Hc(u_i)= 10$ messengers in the up-sector and 
$\sum_i \Hc(q_i) + \Hc(d_i)= 9$ messengers in the down-sector. 
All terms contributing to the determinant are of the same order (which is true in every U(1) model). We can choose the term that is the product of the three diagonal entries. These three entries are then generated from the chain diagrams. For the chains in the up sector we choose (from the total number of possibilities\footnote{In the simple U(1) case, this multiplicity arises from 
permuting the position of the Higgs insertion in each of the three chain diagrams.} that is $7 \times 5 = 35$)
\begin{gather}
\label{u1sol1}
q_1 \! - \! Q_2\! -\! Q_1 \! - \! Q_0 \!- \! U_0 \!- \! U_1 \! - \! U_2 \! - \! u^c_1 \\
q_2 \! -\! Q^\prime_1 \! - \! Q_0^\prime \!- \! U^\prime_0 \!- \! U^\prime_1 \! - \! u^c_2, 
\end{gather}
where the messengers are labelled with their U(1) charges. Note that $y_{33}^U$ is present already on the renormalisable level. For the down sector we choose 
(from $6 \times 4 \times 2 = 48$ possibilities)
\begin{gather}
q_1 \! - \! Q_2\! -\! Q_1 \! - \! Q_0 \!- \! D_0 \!- \! D_1 \! - d^c_1 \\
 q_2 \! -\! Q_1^\prime \! - \! Q^\prime_0 \!- \! D^\prime_0 \!-  \! d^c_2 \\
 q_3 \! - \! D_0^{\prime \prime} \! - \! d^c_3.
 \label{u1sol2}
\end{gather}
In total we have used 10 messengers in the up sector and 9 messengers in the down sector, which is exactly the minimal number required. 

\subsubsection{Higgs UVC}

The charges of the heavy Higgs messengers can be easily inferred from the chain diagram of Fig.~\ref{Hchain}, giving 6 messengers in the up sector and 5 in the down sector which are necessarily distinct. Thus one has in total 11 Higgs messengers (plus conjugates) that we denote by the charge of $h_u$  
\begin{align}
H_u^{(x)} + H_d^{(x)} \quad x= -6,-5,-4,-3,-2,-1,1,2,3,4,5.
\end{align}
The relevant part of the superpotential is given by
\begin{align}
& W \supset~  \sum_x M_x\, H_u^{(x)} H_d^{(x)} + \lambda^{u}_{ij}\, q_i u^c_j\, H_u^{-(\Hc(q_i) + \Hc(u_j))}
+\lambda^{d}_{ij}\, q_i d^c_j\, H_d^{(\Hc(q_i) + \Hc(d_j))} + \nonumber \\
&  \phi \left(\alpha_6 H_d^{(-6)} H_u^{(-5)}+\cdots+ 
\alpha_1 H_d^{(-1)} h_u +\alpha_0 h_d H_u^{(1)}+ \alpha_{-1} H_d^{(1)} H_u^{(2)}
+ \cdots + \alpha_{-4} H_d^{(4)} H_u^{(5)} \right),
\label{Higgs-ex}
\end{align}
where $h_u$ and $h_d$ are the MSSM Higgs fields. Taking $M_x \sim M$, the resulting Yukawa couplings for the up and the down sector are given by
\begin{align}
Y_u& =
\begin{pmatrix}
\lambda^u_{11} \alpha_{(1,6)} \eps^6 &  \lambda^u_{12} \alpha_{(1,5)} \eps^5 & \lambda^u_{13} \alpha_{(1,3)} \eps^3 \\
 \lambda^u_{12} \alpha_{(1,5)} \eps^5 & \lambda^u_{22} \alpha_{(1,4)} \eps^4 &  \lambda^u_{23} \alpha_{(1,2)} \eps^2 \\
 \lambda^u_{13} \alpha_{(1,3)} \eps^3 &   \lambda^u_{23} \alpha_{(1,2)} \eps^2  & \lambda^u_{33}
\end{pmatrix} \\
Y_d & =
\begin{pmatrix}
\lambda^d_{11} \alpha_{(-4,0)} \eps^5 &  \lambda^d_{12} \alpha_{(-3,0)} \eps^4 &  \lambda^d_{13} \alpha_{(-3,0)} \eps^4 \\
\lambda^d_{21} \alpha_{(-3,0)} \eps^4 &  \lambda^d_{22} \alpha_{(-2,0)} \eps^3 &   \lambda^d_{23} \alpha_{(-2,0)} \eps^3 \\
\lambda^d_{31} \alpha_{(-1,0)} \eps^2 &  \lambda^d_{32} \alpha_{(-1,0)} \eps  &  \lambda^d_{33} \alpha_0 \eps
\end{pmatrix},
\label{Yuk-Higgs-ex}
\end{align}
with the shorthand notation
\begin{align}
\alpha_{(X,Y)} \equiv \prod_{x=X, X+1,\ldots, Y} \alpha_x.
\end{align}
Notice that in contrast to the fermion messenger case one does not have to add copies of messengers, since in this case every entry comes with a different coupling $\lambda_{ij}$ which implies full rank.

As explained in Section \ref{HUVC}, Higgs UV completions easily allow for the presence of texture zeros in the Yukawas, simply by removing the corresponding Higgs messenger from the theory (which however sets all Yukawas which the same charge to zero). But the removal of a certain Higgs might interrupt the chain needed for other Yukawa entries. This gap has then to be fixed by using additional singlet messengers. 

As an example, we consider the above U(1) model where we now remove the heavy down-Higgs messenger with charge -4, that is $H_d^{(4)}$, together with its conjugate partner $H_u^{(4)}$. This results in setting the 1-2, 1-3 and 2-1 entries in $Y_d$. to zero.  But at the same time this also removes the coupling $ \phi H_d^{(4)} H_u^{(5)}$ from the theory, so that also the 1-1 entry would be zero as $H_d^{(5)}$ would not take a vev. In order to restore it, one has to introduce singlet messengers. Specifically, one must add vector-like singlets $S_{5,4,3,2} + S_{-5,-4,-3,-2}$ to the theory. 
Besides mass terms, the new allowed couplings are  
\begin{align}
W \supset h_d H_u^{(5)} S_{-5} + \phi \left( S_5 S_{-4} + S_4 S_{-3} + S_3 S_{-2} + S_2 \phi \right).
\end{align}
The last coupling induces a vev for $S_{-2}$, this one a vev for $S_{-3}$, this a vev for $S_{-4}$ and $S_{-5}$, which finally gives together with the first 
interaction a vev to $H_d^{(5)}$.

Let us notice that with a similar procedure one can eliminate the 3-1 entry (by dropping $H_d^{(2)}$), whilst the 2-3 and 3-2 entries cannot be set to zero without removing the 2-2 and the 3-3 entry, respectively. In this U(1) example, one can therefore enforce at most the following texture for $Y_d$:
\begin{equation}
Y_d  \sim
\begin{pmatrix}
\eps^5 &  0 &  0\\
0 &  \eps^3 & \eps^3 \\
0 &  \eps  &  \eps 
\end{pmatrix},
\label{texture}
\end{equation}
that has the interesting feature that the Cabibbo angle arises entirely from the up sector.

\subsection{U(1)$\times$ U(1)$^\prime$}
\label{sec:u1xu1}

Next we consider the U(1)$\times$U(1)$'$ model in Ref.~\cite{MMM1}. The charges of the MSSM superfields are
\begin{align}
 q_{1,2,3} &  : \quad (0,1) \quad (1,0) \quad (0,0)  \nonumber\\
 u^c_{1,2,3} &  : \quad  (0,1) \quad (-1,1) \quad (0,0)  \nonumber\\
 d^c_{1,2,3} &  : \quad  (0,1) \quad (1,0) \quad (1,0), 
\end{align}
the Higgs are neutral and the three flavons $\{ \phi_1$, $\overline{\phi}_1$, $\phi_2 \}$ have charges $\{ (-1,0)$, $(1,0)$, $(0,-1) \}$ and take vevs $\{ \eps_1 \sim \eps^2$, $\overline{\eps}_1 = \eps_1$, $\eps_2 \sim \eps^3\}$. The effective Yukawas are then given by 
\begin{gather}
Y_u \sim 
\begin{pmatrix}
\eps_2^2 & \eps_1 \eps_2^2 & \eps_2 \\
\eps_1 \eps_2 & \eps_2 & \eps_1 \\
\eps_2 & \eps_1 \eps_2 & 1 \\
\end{pmatrix}
\sim 
\begin{pmatrix}
\eps^6 & \eps^8 & \eps^3 \\
\eps^5& \eps^3 & \eps^2\\
 \eps^3 & \eps^5 & 1 \\
\end{pmatrix} 
\nonumber \\
 Y_d  \sim 
\begin{pmatrix}
\eps_2^2 & \eps_1 \eps_2 & \eps_1 \eps_2 \\
\eps_1 \eps_2 & \eps_1^2 & \eps_1^2 \\
\eps_2 & \eps_1  & \eps_1 \\
\end{pmatrix}
\sim 
\begin{pmatrix}
\eps^6 & \eps^5 & \eps^5 \\
\eps^5& \eps^4 & \eps^4\\
\eps^3 & \eps^2 & \eps^2 \\
\end{pmatrix} .
\label{u1xu1yuk}
\end{gather}

\subsubsection{Fermion UVC}

The determinants of the Yukawas in Eq.~(\ref{u1xu1yuk}) are  
\begin{align}
 \det{Y_u} & \sim \eps_2^3 &  \det{Y_d} & \sim \eps_1^3 \eps_2^2,
\end{align}
so that one needs 3 messengers in the up and 5 in the down sector. We again choose in both sectors the diagonal entries, and build the associated chains with right-handed messengers only. In the up sector we take (out of 6 possibilities) the chain
\begin{gather}
q_1 \! - \! U_{(0,-1)} \! - \! U_{(0,0)} \! - \! u_1^c \\
q_2 \! - \! U_{(-1,0)} \! - \! u^c_2,
\end{gather}
while in the down sector we take (out of 18 possibilities)
\begin{gather}
q_1 \! - \! D_{(0,1)} \! - \! D_{(0,0)} \! - \! d_1^c \\
q_2 \! - \! D_{(-1,0)} \! - \! D_{(0,0)}^\prime \! - \! d^c_2\\
q_3 \! - \! D_{(0,0)}^{\prime \prime} \! - \! d^c_3.
\end{gather}
In total we use 3 messengers for the up sector and 5 messengers for the down sector, which is the minimal number needed.  

\subsubsection{Higgs UVC}

By introducing heavy Higgs messengers with charges (plus conjugates):
\begin{align*}
 H_u~: &~ (0,-2),~(1,-2),~(0,-1),~(-1,-1),~(-1,0),~(1,-1) \\
H_d~: &~ (0,-2),~(-1,-1),~(-2,0),~(0,-1),~(-1,0).
\end{align*}
one can generate all entries in the Yukawa matrices of Eq.~(\ref{u1xu1yuk}).

\subsection{SU(3)$_F$}

\label{sec:su3}
Finally we provide an example with a non-abelian flavour symmetry. We take the SU(3) model in Ref~\cite{Ross}.
All the MSSM superfields, $q$, $u^c$ and $d^c$ transform as triplets of SU(3).
Three anti-triplets flavons are introduced with vevs of the form
\begin{equation}
 \frac{\langle\bar{\phi}_3\rangle}{M} \sim (0,0,\eps_3), \quad \frac{\langle\bar{\phi}_{23}\rangle}{M} \sim (0,1,-1)\times \eps_{23},
  \quad \frac{\langle\bar{\phi}_{123}\rangle}{M} \sim (1,1,1)\times \eps_{123},
\end{equation}
with $\eps_{23}=\eps$, $\eps_{123}=\eps^2$ and $\eps_3 = \ord{1}$. The expansion parameter $\eps$ is assumed 
to be different in the up and down sector, with $\eps_u \sim 0.05$, $\eps_d \sim 0.15$. 
In order to differentiate lepton and down-quark mass matrices, a field $\Sigma$ is introduced, with $\langle\Sigma \rangle /M=\sigma \propto (B-L)$.
Unwanted operators are forbidden with additional symmetries, under which the flavons are charged while the MSSM superfields 
are neutral. For our purposes, here we can simply take a single U(1)$_{\Hc}$, with $\Hc(\bar{\phi}_3)=2$,
$\Hc(\bar{\phi}_{23})=1$, $\Hc(\bar{\phi}_{123})=3$, $\Hc(\Sigma)=2$. The MSSM Higgs fields have $\Hc = -4$.
The above set-up gives rise to Yukawas of the form
\begin{equation}
Y_{u,d} \sim 
\begin{pmatrix}
0 & \eps_{123}\eps_{23} & \eps_{123}\eps_{23} \\
\eps_{123}\eps_{23} & \eps_{23}^2 \sigma & \eps_{23}^2 \sigma\\
\eps_{123}\eps_{23} & \eps_{23}^2 \sigma & \eps_3^2 \\
\end{pmatrix}
\sim
\begin{pmatrix}
0 & \eps_{u,d}^3 & \eps_{u,d}^3 \\
\eps_{u,d}^3 & \eps_{u,d}^2 & \eps_{u,d}^2 \\
\eps_{u,d}^3 & \eps_{u,d}^2 & \ord{1} \\
\end{pmatrix}.
\label{su3yukawas}
 \end{equation}

\subsubsection{Fermion UVC}

The determinant of the Yukawas in Eq.~(\ref{su3yukawas}) is 
\begin{equation}
 \det Y_{u,d} \sim \eps^2_3 \eps^2_{23} \eps^2_{123},
\end{equation}
so that 6 messengers in both sectors are required.
From Eq.~(\ref{su3yukawas}), we see that the leading contribution to the determinant is given by the Yukawa entries $Y_{33}$, $Y_{12}$ and $Y_{21}$.
Let us construct a minimal set of messengers for the up sector (the down sector is analogous). We choose to build the three
chains with the Higgs in the middle (in total we have 27 possibilities\footnote{Some of these possibilities might require additional fields in order to write the theory in an SU(3) invariant way: actually only three possibilities have the minimal number 
of fields and are SU(3) invariant.})
\begin{gather}
\label{su3sol1}
q_3 \! - \! Q_{2} \! - \! U_{2} \! - \! u_3^c \\
q_1 \! - \! Q_{3} \! -  \! U_{1} \! -\! u^c_2 \\
q_2 \! - \! Q_{1} \! -  \! U_{3} \! -\! u^c_1,
\label{su3sol2}
\end{gather}
where all 6 messengers are SU(3) singlets and are labelled with the additional $U(1)_{\cal H}$ charge. 

\subsubsection{Higgs UVC}

One needs to introduce (both in the up and in the down sector) a heavy Higgs transforming as $\mathbf{\overline{6}}$ under SU(3) and neutral
under U(1)$_{\Hc}$, which couples directly to the SM fermions.\footnote{A SU(3) triplet messenger $H_{\mathbf{3}}$ would lead to antisymmetric Yukawa matrices.} In addition one needs three anti-triplets Higgs with $\Hc=-3$, $\Hc=-2$, $\Hc=-1$:
\begin{align}
H_{\mathbf{\overline{6}}},~H_{\mathbf{\overline{3}}}^{(-3)},~H_{\mathbf{\overline{3}}}^{(-2)},~H_{\mathbf{\overline{3}}}^{(-1)},~
\label{su3uvc1}
\end{align}
or alternatively two sexplet singlets with $\Hc= 2$, $\Hc= 4$: 
\begin{align}
\label{SU3b}
 H_{\mathbf{\overline{6}}},~S_{\mathbf{\overline{6}}}^{(2)},~S_{\mathbf{\overline{6}}}^{(4)},
\end{align}
plus the corresponding conjugate fields.

\section{Explicit examples of RG induced flavour violation}
\label{RGE-appendix}
We now show an explicit example of the RG effect discussed in Section~\ref{pheno}.
We consider the U(1) model described in Appendix~\ref{sec:u1} in the case of HUVC. 
The starting point is the superpotential in Eq.~(\ref{Higgs-ex}). As an effect of the flavour symmetry breaking, Higgs messengers mix among themselves. 
The RGEs for the soft masses can be conveniently derived in the messenger mass eigenbasis. Concentrating on the down 1-2 sector, we obtain
for the mass splittings in leading log approximation:
\begin{align}
 (m^2_{\tilde d})_{22} - (m^2_{\tilde d})_{11}  & \approx \frac{12}{16 \pi^2} \tilde{m}_0^2 \left[ (\lambda^{d\dag} \lambda^d)_{11} -
 (\lambda^{d\dag} \lambda^d)_{22}\right] \log\frac{M_S}{M} , \\
(m^2_{\tilde q})_{22} - (m^2_{\tilde q})_{11}  & \approx \frac{6}{16 \pi^2} \tilde{m}_0^2 \left[ (\lambda^{u\dag} \lambda^u)_{11} -
 (\lambda^{u\dag} \lambda^u)_{22}+ (\lambda^{d\dag} \lambda^d)_{11} -
 (\lambda^{d\dag} \lambda^d)_{22}\right]\log\frac{M_S}{M} ,
\end{align}
where we have neglected the contribution of the A-terms.
These splittings induce the following contributions to the off-diagonal entries in the super-CKM basis:
\begin{align}
\label{sckm1}
 (m^2_{\tilde d})^{\rm ROT}_{12} &= \left[ (m^2_{\tilde d})_{22} - (m^2_{\tilde d})_{11}\right] \theta_{12}^{D R}, \\
 (m^2_{\tilde q})^{\rm ROT}_{12} &= \left[ (m^2_{\tilde q})_{22} - (m^2_{\tilde q})_{11} \right] \theta_{12}^{D L}.
\label{sckm2}
\end{align}
The mixing angles can be easily estimated from Eq.~(\ref{Yuk-Higgs-ex}):
\begin{equation}
 \theta_{12}^{D R} \approx \frac{\lambda^d_{21}}{\lambda^d_{22}} \alpha_{-3} ~\eps\,,  \quad\quad 
\theta_{12}^{D L} \approx \frac{\lambda^d_{12}}{\lambda^d_{22}} \alpha_{-3} ~\eps\,.
\end{equation}
As mentioned above there is also a contribution to the off-diagonal entries directly generated by the running, that at leading log reads:
\begin{align}
 (m^2_{\tilde d})^{\rm RG}_{12} \approx \frac{12}{16 \pi^2} \tilde{m}_0^2 &\left(\lambda^{d *}_{11}\lambda^d_{12} \alpha_{-4}^* 
+ \lambda^{d *}_{21}\lambda^d_{22} \alpha_{-3}^*   +\lambda^{d *}_{31}\lambda^d_{32} \alpha_{-1}^* \right) \eps \,\log\frac{M_S}{M} \,,\\
 (m^2_{\tilde q})^{\rm RG}_{12} \approx \frac{6}{16 \pi^2} \tilde{m}_0^2 &\left(\lambda^{d *}_{11}\lambda^d_{21} \alpha_{-4}^* 
+ \lambda^{d *}_{12}\lambda^d_{22} \alpha_{-3}^*   +\lambda^{d *}_{13}\lambda^d_{23} \alpha_{-1}^*  \right. \nonumber\\
& \left. + \lambda^{u *}_{11}\lambda^d_{21} \alpha_{6}^* 
+ \lambda^{u *}_{12}\lambda^d_{22} \alpha_{5}^*   +\lambda^{u *}_{13}\lambda^d_{23} \alpha_{3}^*  \right) \eps \,\log\frac{M_S}{M} \,,
\end{align}
where again the effect of the trilinears is neglected.
As we can see, these expressions do not cancel in general against the contributions of Eqs.~(\ref{sckm1},~\ref{sckm2}), since they involve different
$\ord{1}$ coefficients. They also show that the enhancement factor $\sim$10 we considered for our estimates is a reasonable approximation.

Let us now discuss the same effect for the SU(3) model with HUVC. We start from the following superpotential for the UVC of 
Eq.~(\ref{su3uvc1}):
\begin{align}
W & \supset \lambda_d q_i d^c_j H_{\mathbf{\overline{6}}}^{(ij)} + \alpha_3 \bar{H}_{\mathbf{\overline{6}}}^{(ij)} \phi_3^i H_{\mathbf{\overline{3}}}^{(-2) j}
+ \alpha_{23} \bar{H}_{\mathbf{\overline{6}}}^{(ij)} \phi_{23}^i H_{\mathbf{\overline{3}}}^{(-1) j} + \ldots
\label{su3W}
\end{align}
where $i$ and $j$ denote SU(3)$_F$ indices and we omitted the messenger mass terms and other couplings that are not relevant for the present discussion. 
From the mass eigenbasis for the Higgs messengers, we see that $H_{\mathbf{\overline{6}}}$ acquires small components of the other messengers:
\begin{align}
H_{\mathbf{\overline{6}}}^\prime \sim H_{\mathbf{\overline{6}}} + \alpha_3 \frac{\langle\phi_3\rangle}{M} H_{\mathbf{\overline{3}}}^{(-2)}
+ \alpha_{23} \frac{\langle\phi_{23}\rangle}{M} H_{\mathbf{\overline{3}}}^{(-1)} + \ldots
\end{align}
Plugging this expression in Eq.~(\ref{su3W}), we see that $q_i d^c_j$ couple with $H_{\mathbf{\overline{3}}}^{(-2)}$ and  $H_{\mathbf{\overline{3}}}^{(-1)}$ non-universal in flavour, as $\langle\phi_{3}^i\rangle = 0$ for $i \neq 3$ and
$\langle\phi_{23}^i\rangle = 0$ for $i \neq 2,3$. In the running for $m^2_{\tilde q}$ and $m^2_{\tilde d}$, this induces a splitting
$\propto \lambda_d^2 \alpha_{23}^2 \eps_{23}^2 \sim \eps_d^2$ between the first and second generation sfermion masses and a splitting
$\propto \lambda_d^2 \alpha_{3}^2 \eps_{3}^2 \sim \ord{1}$ between the third generation mass and the other two. Therefore
the induced off-diagonal entries $(m^2_{\tilde q})_{12}$ and $(m^2_{\tilde d})_{12}$ are just suppressed by an additional 
$\eps_d^2$ factor with respect to the abelian case discussed above and the entries involving the third family have no further suppression.


\begin{thebibliography}{99}

\bibitem{FN}
  C.~D.~Froggatt and H.~B.~Nielsen,
  Nucl.\ Phys.\  B {\bf 147} (1979) 277.

\bibitem{MMM1}
M.~Leurer, Y.~Nir, N.~Seiberg,
  Nucl.\ Phys.\  {\bf B398 } (1993)  319-342
  [hep-ph/9212278].
  
\bibitem{MMM2}
 M.~Leurer, Y.~Nir, N.~Seiberg,
  Nucl.\ Phys.\  {\bf B420}, 468-504 (1994)
  [hep-ph/9310320].

\bibitem{quilt}
  P.~Ramond, R.~G.~Roberts and G.~G.~Ross,
  Nucl.\ Phys.\  B {\bf 406} (1993) 19
  [arXiv:hep-ph/9303320].
 
\bibitem{Kadota:2010cz}
  K.~Kadota, J.~Kersten and L.~Velasco-Sevilla,
  Phys.\ Rev.\ D {\bf 82} (2010) 085022
  [arXiv:1007.1532 [hep-ph]].

\bibitem{Luca}
  I.~de Medeiros Varzielas and L.~Merlo,
  JHEP {\bf 1102} (2011) 062
  [arXiv:1011.6662 [hep-ph]].

\bibitem{Antusch:2010es}
  S.~Antusch, S.~F.~King and M.~Spinrath,
  Phys.\ Rev.\ D {\bf 83} (2011) 013005
  [arXiv:1005.0708 [hep-ph]].
  
\bibitem{CLPZlow}
  L.~Calibbi, Z.~Lalak, S.~Pokorski and R.~Ziegler,
  arXiv:1204.1275 [hep-ph].

 \bibitem{LPR}
 For a recent discussion see
  Z.~Lalak, S.~Pokorski, G.~G.~Ross,
  JHEP {\bf 1008}, 129 (2010)
  [arXiv:1006.2375 [hep-ph]].

\bibitem{MFV}
  G.~D'Ambrosio, G.~F.~Giudice, G.~Isidori and A.~Strumia,
  Nucl.\ Phys.\ B {\bf 645} (2002) 155
  [hep-ph/0207036].

\bibitem{typeII}
See e.g.~A.~Rossi,
  Phys.\ Rev.\ D {\bf 66} (2002) 075003
  [hep-ph/0207006].
  

  \bibitem{Magic}
  L.~Calibbi, L.~Ferretti, A.~Romanino, R.~Ziegler,
  Phys.\ Lett.\  {\bf B672}, 152-157 (2009)
  [arXiv:0812.0342 [hep-ph]].


\bibitem{Buras-Paride}
  For a recent discussion, see 
  W.~Altmannshofer, A.~J.~Buras, S.~Gori, P.~Paradisi and D.~M.~Straub,
  Nucl.\ Phys.\ B {\bf 830} (2010) 17
  [arXiv:0909.1333 [hep-ph]].


\bibitem{Conseq} 
  L.~Calibbi, L.~Ferretti, A.~Romanino and R.~Ziegler,
  JHEP {\bf 0903}, 031 (2009)
  [arXiv:0812.0087 [hep-ph]].


\bibitem{models-su3}
  S.~F.~King and G.~G.~Ross,
  Phys.\ Lett.\ B {\bf 520} (2001) 243
  [hep-ph/0108112];
  G.~G.~Ross, L.~Velasco-Sevilla and O.~Vives,
  Nucl.\ Phys.\ B {\bf 692} (2004) 50
  [hep-ph/0401064].


\bibitem{models-u2}
  A.~Pomarol and D.~Tommasini,
  Nucl.\ Phys.\ B {\bf 466} (1996) 3
  [hep-ph/9507462];
  R.~Barbieri, G.~R.~Dvali and L.~J.~Hall,
  Phys.\ Lett.\ B {\bf 377} (1996) 76
  [hep-ph/9512388].


 \bibitem{Ross}
   I.~de Medeiros Varzielas, G.~G.~Ross,
  Nucl.\ Phys.\  {\bf B733}, 31-47 (2006)
  [hep-ph/0507176].



\bibitem{Borzumati:1986qx}
  F.~Borzumati, A.~Masiero,
  Phys.\ Rev.\ Lett.\  {\bf 57 } (1986)  961.
 
\bibitem{Barbieri-Hall}
  R.~Barbieri and L.~J.~Hall,
  Phys.\ Lett.\ B {\bf 338} (1994) 212
  [hep-ph/9408406];
  R.~Barbieri, L.~J.~Hall and A.~Strumia,
  Nucl.\ Phys.\ B {\bf 445} (1995) 219
  [hep-ph/9501334].


  \bibitem{models2}
  G.~Altarelli, F.~Feruglio and C.~Hagedorn,
  JHEP {\bf 0803} (2008) 052
  [arXiv:0802.0090 [hep-ph]];
  F.~Bazzocchi, L.~Merlo and S.~Morisi,
  Nucl.\ Phys.\ B {\bf 816} (2009) 204
  [arXiv:0901.2086 [hep-ph]].









\bibitem{Masiero-Vives-Vempati}
  A.~Masiero, S.~K.~Vempati and O.~Vives,
  arXiv:0711.2903 [hep-ph].

\bibitem{Ciuchini:2007cw}
  M.~Ciuchini, E.~Franco, D.~Guadagnoli, V.~Lubicz, M.~Pierini, V.~Porretti and L.~Silvestrini,
  Phys.\ Lett.\ B {\bf 655} (2007) 162
  [hep-ph/0703204].


\bibitem{GinoNir}
  G.~Isidori, Y.~Nir and G.~Perez,
  Ann.\ Rev.\ Nucl.\ Part.\ Sci.\  {\bf 60} (2010) 355
  [arXiv:1002.0900 [hep-ph]].

\bibitem{MEG}
  J.~Adam {\it et al.}  [MEG Collaboration],
  Phys.\ Rev.\ Lett.\  {\bf 107} (2011) 171801
  [arXiv:1107.5547 [hep-ex]].


\bibitem{Paride}
  M.~Ciuchini, A.~Masiero, P.~Paradisi, L.~Silvestrini, S.~K.~Vempati and O.~Vives,
  Nucl.\ Phys.\ B {\bf 783} (2007) 112
  [hep-ph/0702144 [HEP-PH]].


\bibitem{su3flavour}
  L.~Calibbi, J.~Jones-Perez and O.~Vives,
  Phys.\ Rev.\ D {\bf 78} (2008) 075007
  [arXiv:0804.4620 [hep-ph]];
  L.~Calibbi, J.~Jones-Perez, A.~Masiero, J.~-h.~Park, W.~Porod and O.~Vives,
  Nucl.\ Phys.\ B {\bf 831} (2010) 26
  [arXiv:0907.4069 [hep-ph]].


\bibitem{Jaxodraw}
  D.~Binosi, J.~Collins, C.~Kaufhold and L.~Theussl,
  Comput.\ Phys.\ Commun.\  {\bf 180} (2009) 1709
  [arXiv:0811.4113 [hep-ph]];
  D.~Binosi and L.~Theussl,
  Comput.\ Phys.\ Commun.\  {\bf 161} (2004) 76
  [hep-ph/0309015].



 \bibitem{CKLP} 
 P.~H.~Chankowski, K.~Kowalska, S.~Lavignac and S.~Pokorski,
  Phys.\ Rev.\  D {\bf 71} (2005) 055004
  [arXiv:hep-ph/0501071]. 
  

\end{thebibliography}
\end{document}